# Digital Fourier transform spectroscopy:
# a high-performance, scalable technology for on-chip spectrum analysis

**Derek M. Kita[1,*], Brando Miranda[2], David Favela[3], David Bono[1], Jérôme Michon[1], Hongtao Lin[1], Tian Gu[1] and Juejun Hu[1,*]**

[1]*Department of Materials Science & Engineering, Massachusetts Institute of Technology, Cambridge, Massachusetts, USA*
[2]*Center for Brains, Minds & Machines, Massachusetts Institute of Technology, Cambridge, Massachusetts, USA*
[3]*Department of Mechanical Engineering, Massachusetts Institute of Technology, Cambridge, Massachusetts, USA*

*dkita@mit.edu, hujuejun@mit.edu

**Introductory paragraph**

Optical spectrum analysis is the cornerstone of spectroscopic sensing, optical network performance monitoring, and hyperspectral imaging. While conventional high-performance spectrometers used to perform such analysis are often large benchtop instruments, on-chip spectrometers have recently emerged as a promising alternative with apparent Size, Weight, and Power (SWaP) advantages. Existing on-chip spectrometer designs, however, are limited in spectral channel count and signal-to-noise ratio (SNR). Here we demonstrate a transformative on-chip digital Fourier transform (dFT) spectrometer that can acquire high-resolution spectra via time-domain modulation of a reconfigurable Mach-Zehnder interferometer. The device, fabricated and packaged using industry-standard silicon photonics technology, claims the multiplex advantage to dramatically boost SNR and unprecedented scalability capable of addressing exponentially increasing numbers of spectral channels. We further implemented machine learning regularization techniques to spectrum reconstruction and achieved significant noise suppression and spectral resolution enhancement beyond the classical Rayleigh criterion.

Optical spectrometers are extensively applied to sensing, materials analysis, and optical network monitoring. Conventional spectrometers are bulky instruments often involving mechanical moving parts, which severely compromises their deployment versatility and increases cost. Photonic integration offers a solution to miniaturize spectrometers into a chip-scale platform, albeit often at the cost of performance and scalability. Existing on-chip spectrometers mostly rely on dispersive elements such as gratings[1-6], holograms[7,8], and microresonators[9-11]. These devices suffer serious SNR penalties when designed for high spectral resolution, as a result of spreading input light over many spectral channels. Moreover, the device footprint and complexity scale *linearly* with the spectral channel number *N*, as each channel requires an individually addressed receiver (photodetector) and the spectral resolution is inversely proportional to the optical path length (OPL). The SNR degradation and linear scaling behavior preclude high-performance on-chip spectrometers with channel counts rivaling their benchtop counterparts, which typically have hundreds to thousands of spectral channels. We note that the constraints also apply to spectrometers based on the wavelength multiplexing principle, where each receiver captures an ensemble of monochromatic light rather than one single wavelength[12-18]*.

Unlike dispersive spectrometers, Fourier transform infrared (FTIR) spectrometers overcome the trade-off between SNR and spectral resolution benefiting from the multiplex advantage, also known as the Fellgett's advantage[19]. Traditional benchtop FTIR spectrometers use moving mirrors to generate a tunable OPL, a design not readily amenable to planar photonic integration. On-chip FTIR spectrometers instead rely on thermo-optic or electro-optic modulation to change the OPL in a waveguide[20-23]. The miniscule refractive index modifications produced by these effects, however, result in large device size and constrain the practically attainable spectral resolution to tens of cm$^{-1}$ in wave number, far inferior compared to their benchtop counterparts.

In this letter, we propose and experimentally demonstrate a novel spectrometer architecture that resolves the performance and scalability challenges. The centerpiece of the spectrometer is a reconfigurable Mach-Zehnder interferometer (MZI) illustrated in Fig. 1a. Each arm consists of *j*/2 cascaded sets of optical switches connected by waveguides of varying lengths, where *j* is an even integer. When light propagates through the reference paths (marked with black color) in both MZI arms, the MZI is balanced with zero OPL difference between the two arms. Lengths of the waveguide paths in red differ from the reference paths by a power of two times Δ*L*. Each permutation of the switches thus corresponds to a unique OPL difference between the arms, covering 0 to $(2^j - 1) \cdot n_g \cdot \Delta L$ with a step size of $n_g \cdot \Delta L$, where $n_g$ represents the waveguide group index. Unlike traditional FTIR spectrometers where the OPL is continuously tuned, resembling an analog signal, our digital Fourier transform (dFT) spectrometer derives its name from the set of "digitized" binary optical switches, with the state of each corresponding to a unique permutation of the spectrometer and a unique OPL difference. The number of spectral channels, defined by the distinctive optical states the device furnishes, is:

$$N = 2^j, \qquad (1)$$

and the spectral resolution is given following the Rayleigh criterion[24-26]:

$$\delta\lambda = \frac{\lambda^2}{(2^j - 1) \cdot n_g \Delta L} \approx \frac{1}{2^j} \cdot \frac{\lambda^2}{n_g \Delta L}, \qquad (2)$$

where λ denotes the center wavelength. The equations reveal three key advantages of the dFT spectrometer technology over state-of-the-art. First, both the spectral channel count and resolution scale *exponentially* with the number of cascaded switch stages. This unique exponential scaling behavior allows high-resolution spectroscopy with a radically simplified device architecture. Second, direct modification of the waveguide path offers over 100 times larger OPL tuning compared to thermo-optic or electro-optic-based index modulation, enabling superior spectral resolution within a compact device. An additional benefit is that the device is far less sensitive to temperature variations than existing on-chip FTIR spectrometers, as the temperature-induced OPL fluctuation scales linearly with the physical length of the interferometer arms. Third, the device benefits from the multiplex advantage to ensure significantly enhanced SNR over the dispersive devices. Moreover, the spectrometer only requires a single-element photodetector rather than a linear array, which further reduces system complexity and cost.

---
* For multiple-scattering-based spectrometers, the device dimension scales with spectral resolution quadratically.

We experimentally validated the dFT spectrometer concept by demonstrating a 64-channel device ($j = 6$) operating at the telecommunication C-band. The device was fabricated leveraging a commercial silicon photonics foundry process, where the optical switches employ a custom compact thermo-optic phase shifter design[27]. Figure 1b presents a micrograph of the spectrometer after front-end-of-line silicon fabrication. The chip was subsequently packaged with bonded fiber arrays and electrical connectors. Details of the fabrication and packaging processes are elaborated in the Methods section. The spectrometer also integrates an on-chip germanium photodetector and a standard FC/PC fiber connector interface, making it a standalone "plug-and-play" device for optical spectrum analysis (Fig. 1c).

The spectrometer was characterized using a setup depicted in Fig. 2a. High-resolution transmittance spectra of the device were first recorded by wavelength sweeping a tunable laser between 1550 and 1570 nm, for all 64 permutations of the switch on/off combinations. The 64 spectra are plotted in Fig. 2c, each associated with a unique OPL difference between the MZI arms.

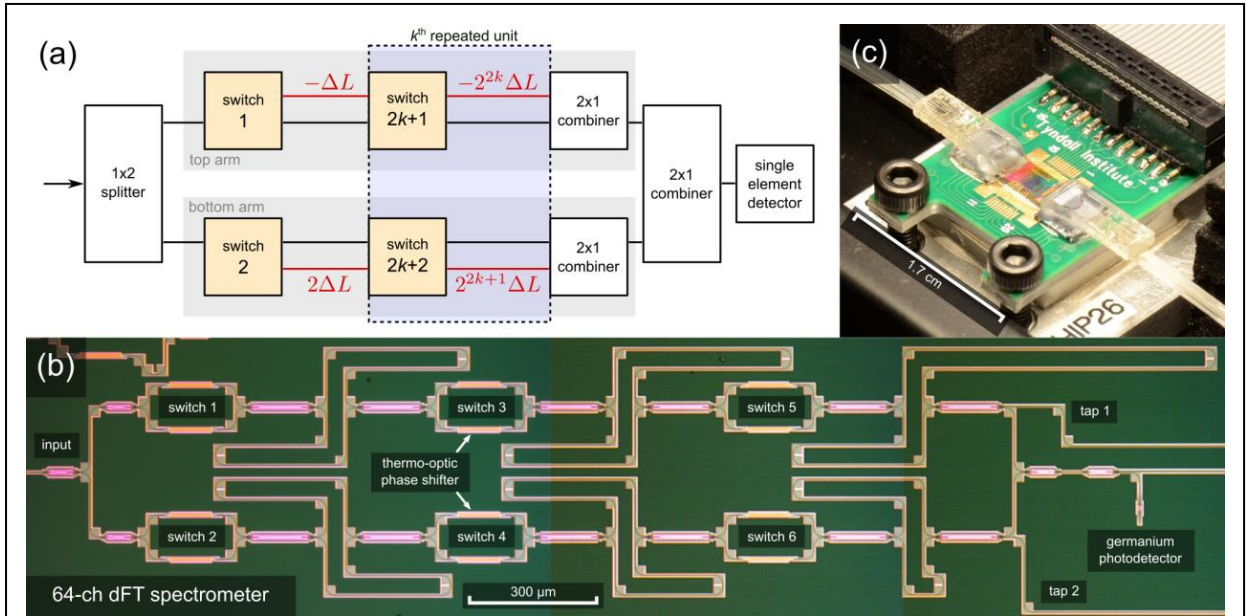

**Fig. 1.** (a) Block diagram illustrating the generic structure of a dFT spectrometer with $j$ switches and $K = j/2 - 1$ repeated stages indexed by $k \in [1, K]$; (b) top-view optical micrograph of the 64-channel dFT spectrometer after front-end-of-line silicon fabrication, showing the interferometer layout, the thermo-optic switches and waveguide-integrated germanium photodetector; (c) photo of the fully-packaged, "plug-and-play" dFT spectrometer with standard FC/PC fiber interface and a ribbon cable for control and signal read-out.

The ensemble of spectra forms an $m \times n$ calibration matrix $\mathbf{A}$. Each row of $\mathbf{A}$ represents a transmittance spectrum and contains $n = 801$ elements, the number of wavelength points in the scan. Each column corresponds to a discretely sampled interferogram of the narrow-band laser and contains $m = 64$ elements. The intensity measured by the detector for an arbitrary input polychromatic signal (represented by a column vector $\mathbf{x}$ with 801 elements) is:

$$\mathbf{y} = \mathbf{A}\mathbf{x}, \tag{3}$$

where the interferogram $\mathbf{y}$ is a column vector with 64 elements, each gives the detector output at a particular switch permutation. The vector $\mathbf{y}$ was measured by recording the detector output at all 64 permutation states. Since we measured $\mathbf{y}$ with size 64 to infer $\mathbf{x}$ with size 801, the system is underconstrained and therefore regularization techniques are required to specify a unique solution.

To do this, we applied a non-negative elastic net method[28] with a smoothing prior on the first derivative. This method, that we call "elastic-$D_1$" from here on, solves the regularization problem:

$$\min_{\mathbf{x},\mathbf{x}>0}\left\{\|\mathbf{y}-\mathbf{A}\mathbf{x}\|_2^2 + \alpha_1\|\mathbf{x}\|_1 + \alpha_2\|\mathbf{x}\|_2^2 + \alpha_3\|\mathbf{D}_I\mathbf{x}\|_2^2\right\}. \tag{4}$$

where $\alpha_1$, $\alpha_2$, and $\alpha_3$ are hyperparameters that weight the corresponding $L_1$- and $L_2$-norms on $\mathbf{x}$, and the $L_2$-norm on the first derivative specified by the matrix $\mathbf{D}_1$. The combination of bounds on the $L_1$-norm (that induces sparsity on the spectrum), $L_2$-norm (that bounds magnitude of the spectrum), and first derivative of the spectrum (that sets the desired smoothing) produce exceptional reconstructions on both broad and narrow spectral features without requiring knowledge of the true input spectrum. Since Eq. 4 is a non-negative quadratic program, it is readily solvable with standard convex optimization tools[29]. Lastly, we determine optimal values of the hyperparameters through leave-one-out cross-validation[30], through which we take two consecutive measurements of the interferogram ($\mathbf{y}_1$ and $\mathbf{y}_2$) and choose hyperparameters that maximize the coefficient of determination $R^2(\mathbf{A}_2\mathbf{x}, \mathbf{y}_2)$, where $\mathbf{A}_1$ and $\mathbf{A}_2$ are two separate measurements of the basis, and $x$ is the computed spectrum from $y_1$ and $A_1$. Details of the algorithm are presented in the Supplementary Information.

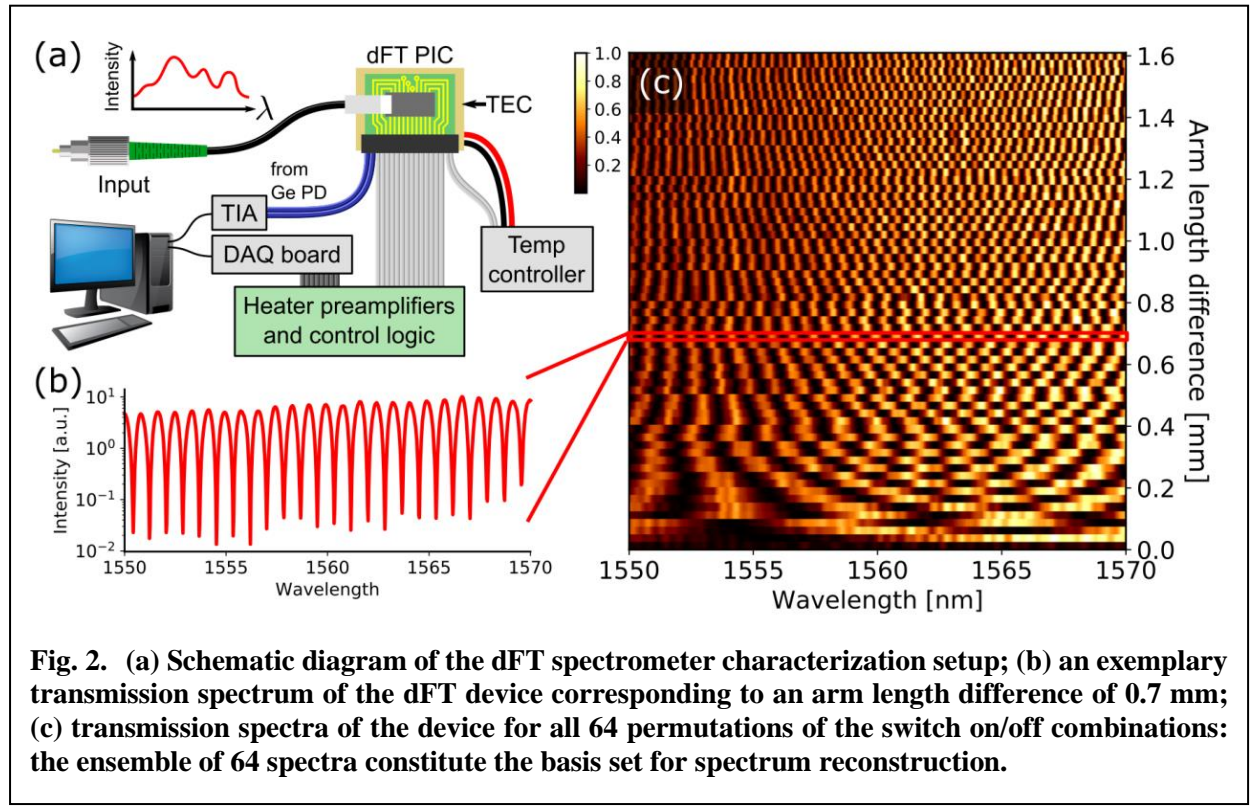

**Fig. 2.** (a) Schematic diagram of the dFT spectrometer characterization setup; (b) an exemplary transmission spectrum of the dFT device corresponding to an arm length difference of 0.7 mm; (c) transmission spectra of the device for all 64 permutations of the switch on/off combinations: the ensemble of 64 spectra constitute the basis set for spectrum reconstruction.

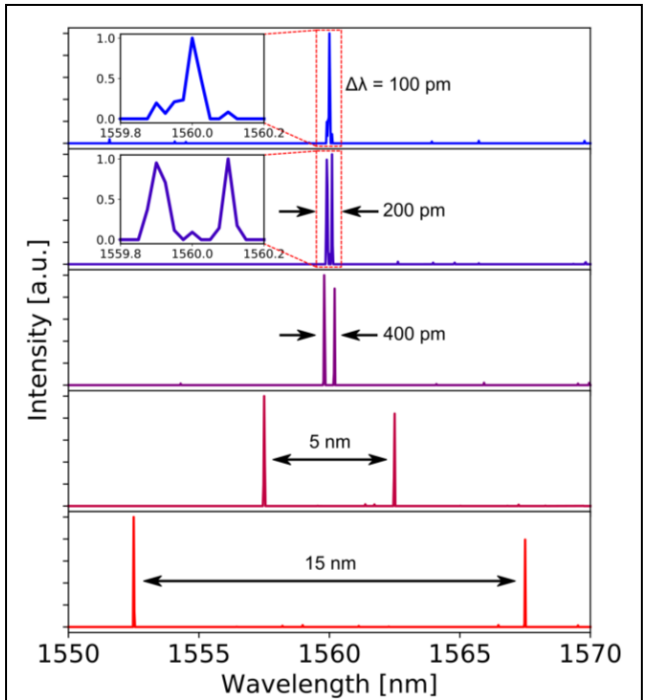

**Fig. 3.** Spectra consisting of two laser lines with varying spacing measured using the 64-channel dFT spectrometer and reconstructed by applying the elastic-$D_1$ algorithm. Inset shows zoomed-in images of the narrow spectral features for input laser lines with 100 pm and 200 pm spacing.

To demonstrate the versatility of the dFT spectrometer, we applied the elastic-$D_1$ reconstruction technique to experimentally measured interferograms for two types of polychromatic inputs, sparse signals consisting of discrete laser lines and broadband signals with complex spectral features (see Methods). Figure 3 plots the reconstructed spectra comprising two laser lines with slightly different amplitudes and varying wavelength spacing. The elastic-$D_1$ technique precisely reproduces the laser wavelengths with ± 0.025 nm accuracy, only limited by the finite wavelength step size of the calibration matrix (0.025 nm). The spectral resolution of our device, determined here by the minimum resolvable wavelength detuning between two laser lines, significantly outperforms the Rayleigh criterion of 0.4 nm with an experimentally determined value of 0.2 nm. The enhanced reconstruction quality is a result of the elastic-$D_1$ method's automatic consideration of the tradeoffs between spectral sparsity, magnitude, and smoothness (Supplementary Information). Figure 4 compares the spectra of three unique broadband inputs recorded using a benchtop optical spectrum analyzer as a reference and reconstructed spectra using the same 64-channel dFT device and elastic-$D_1$ algorithm. The high reconstruction quality on arbitrary input spectra with characteristic spectral features ranging from several nanometers to well below the Rayleigh limit validates dFT spectroscopy as a generic, powerful tool for quantitative spectroscopy.

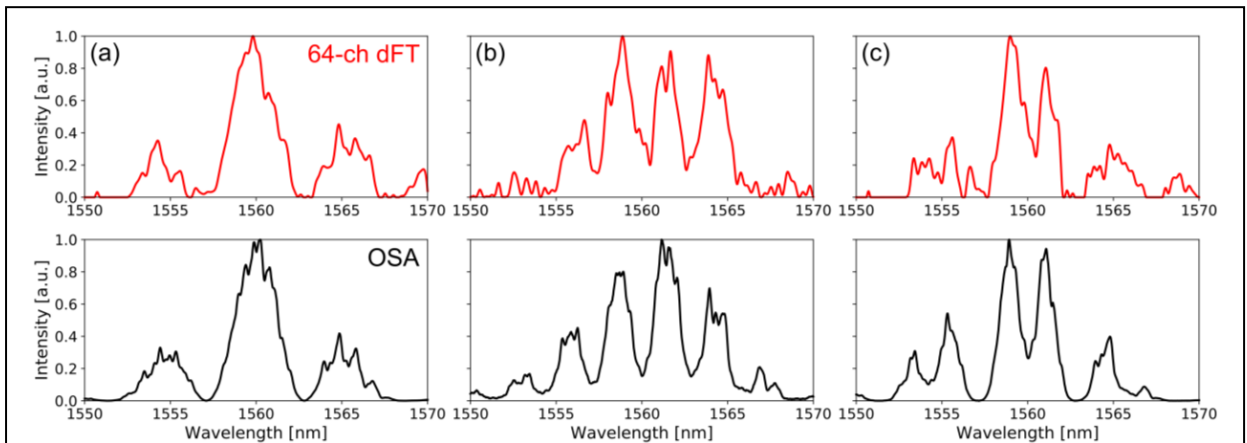

**Fig. 4.** Examples of broadband, arbitrary signal reconstruction showing spectra measured by a benchtop optical spectrum analyzer (black curves) and the dFT spectrometer (red curves).

In conclusion, this work pioneers dFT spectroscopy as a high-performance, scalable solution for on-chip optical spectrum analysis. Its unique exponential scalability in performance, superior SNR leveraging the multiplex advantage, reduced thermal sensitivity, as well as compact and remarkably simplified device design are among the key advantages of the technology. Moreover, its proven compatibility with industry-standard foundry processes enables scalable manufacturing and drastic cost reduction. We further implemented machine learning regularization techniques to achieve significant noise suppression and resolution enhancement. The powerful combination of dFT spectroscopy and machine learning techniques will empower future applications of spectroscopy such as chem/bio sensors-on-a-chip, space-borne spectroscopy, and optical network monitoring.

## Methods

**Device fabrication.** Device layout and mask generation was done using Luceda's IPKISS design framework and tools. The dFT spectrometer chips were fabricated on the imec-ePIXfab active silicon photonics platform (ISIPP25G) multi-project wafer service. Standard passive and active components (excluding the custom compact thermo-optic phase modulator) from the imec-ePIXfab process design kit (PDK) library were used to construct the dFT spectrometer. The devices were subsequently packaged at Tyndall National Institute with fiber grating coupler arrays and electrical connections on a thermoelectric cooler for temperature control.

**Device characterization.** Packaged devices were characterized at MIT with a swept single-frequency external cavity laser to determine the wavelength response of the device for each switch state, using the integrated germanium detector for signal readout. The thermo-optic switches were initially calibrated by tuning the heater powers until the frequency response of the top arm or bottom arm was flat, as measured from one of two "tap" ports, as indicated in Fig. 1. Heater preamplifiers, programmable digital control logic, and photodetector transimpedance amplifiers with variable gain are all implemented with custom electronics and automation software. The elastic-$D_1$ technique was implemented in Python with free software for convex optimization[29]. Testing on two input-lasers with different wavelength spacing was performed by combining two separate tunable continuous-wave external cavity lasers through a $2 \times 2$ beam splitter with one output port to an optical spectrum analyzer and the second to the dFT spectrometer. To generate the broadband input signal shown in Figs. 4a and 4b, we couple the amplified spontaneous emission from an erbium-doped fiber amplifier (EDFA) to on-chip imbalanced MZI structures with arm length differences of 100 μm and 200 μm, respectively. The spectrum in Fig. 4c is obtained by passing light through both MZI structures connected in series. The complex spectral features are attributed to both interference in the imbalanced MZI as well as Fabry-Perot fringes due to multiple reflections at connectors and couplers. The reference spectra were recorded using a Yokogawa AQ6375B optical spectrum analyzer. Raw data from the basis measurements and all interferogram measurements are available from the corresponding author upon request.

**Code availability.** Custom code written in Python to perform spectral reconstruction via the elastic-$D_1$ regularized regression technique is available upon request.

## Acknowledgments

The authors gratefully thank Lionel C. Kimerling, Anu Agarwal, and Rajeev Ram for providing access to device measurement facilities. Funding support is provided by the National Science Foundation under award number 1709212, MIT SENSE.nano Seed Grant, and the Department of Energy under Grant DE-NA0002509. D.K. acknowledges the Kavanagh Fellowship for Technology Commercialization provided by the Saks Kavanaugh Foundation for financial support.

## Author contributions

D.K. designed and characterized the spectrometer device. B.M. and D.K. developed the machine learning algorithms for spectrum reconstruction. D.F., D.B., and D.K. designed and assembled the electronics for device testing. J.M. and H.L. assisted in device characterization. J.H. conceived the spectrometer concept and supervised the research. T.G. and D.K. contributed to the concept formulation. All authors contributed to technical discussions and writing the paper.

## Competing financial interests

The authors declare no competing financial interests.